\documentclass[amsmath,amssymb,twocolumn,showpacs,superscriptaddress,prl,longbibliography]{revtex4-1}

\usepackage{dcolumn}
\usepackage{bm}
\usepackage{graphicx}
\usepackage{epstopdf}
\usepackage{wrapfig}
\usepackage{hyperref}
\usepackage{array} 
\usepackage{listings}
\usepackage[para,online,flushleft]{threeparttablex}
\usepackage{booktabs,dcolumn}
\usepackage{filecontents}
\usepackage{xcolor}
\usepackage{siunitx}
\usepackage{bibunits}
\usepackage{lipsum}
\usepackage{upgreek} 

\usepackage{etoolbox}
\usepackage{comment}
\usepackage{soul}
\makeatletter
\newcommand*{\newbibstartnumber}[1]{%
  \apptocmd{\thebibliography}{%
    \global\c@NAT@ctr #1\relax
    \addtocounter{NAT@ctr}{-1}%
  }{}{}%
}
\newcommand\titlelowercase[1]{\texorpdfstring{\lowercase{#1}}{#1}}
\makeatother

\begin{document}

\preprint{APS/123-QED}

\title{Precision Spectroscopy of Fast, Hot Exotic Isotopes Using Machine Learning Assisted
Event-by-Event Doppler Correction}

\author{S.~M.~Udrescu} 
\email{sudrescu@mit.edu}
\affiliation{Massachusetts Institute of Technology,~Cambridge,
~MA~02139,~USA}
\author{D.~A.~Torres}
\affiliation{Departamento de F\'{\i}sica, Universidad ~Nacional ~de ~Colombia,~Bogot\'a,~Colombia.}
\author{R.~F.~Garcia Ruiz}
\affiliation{Massachusetts Institute of Technology,~Cambridge,
~MA~02139,~USA}
\date{\today}

\begin{abstract}

We propose an experimental scheme for performing sensitive, high-precision laser spectroscopy studies on fast exotic isotopes. By inducing a step-wise resonant ionization of the atoms travelling inside an electric field and subsequently detecting the ion and the corresponding electron, time- and position-sensitive measurements of the resulting particles can be performed. Using a Mixture Density Network (MDN), we can leverage this information to predict the initial energy of individual atoms and thus apply a Doppler correction of the observed transition frequencies on an event-by-event basis. We conduct numerical simulations of the proposed experimental scheme and show that kHz-level uncertainties can be achieved for ion beams produced at extreme temperatures ($> 10^8$ K), with energy spreads as large as $10$ keV and non-uniform velocity distributions. The ability to perform in-flight spectroscopy, directly on highly energetic beams, offers unique opportunities to studying short-lived isotopes with lifetimes in the millisecond range and below, produced in low quantities, in hot and highly contaminated environments, without the need for cooling techniques. Such species are of marked interest for nuclear structure, astrophysics, and new physics searches.

\end{abstract}

\maketitle

\section{Introduction}
With the advent of new radioactive beam facilities worldwide, short-lived nuclei that hitherto have only existed in stellar explosions, are being created artificially in the laboratory, extending our exploration of the nuclear chart to extreme proton-to-neutron ratios \cite{arrowsmith2023,Yan23,block2021recent}. These new unstable isotopes, typically with lifetimes of just a fraction of a second, are critical for our fundamental understanding of nuclei and nuclear matter \cite{Yan23}. The major challenge of  current experimental nuclear physics is to develop sensitive and precise techniques to enable the study of these exotic isotopes, commonly produced at high temperature and with yields of just a few isotopes per second \cite{arrowsmith2023,Yan23,block2021recent}.   
    
Laser spectroscopy has long been established as an important tool for studying the properties of unstable nuclei \cite{Yan23}. This technique allows the extraction of nuclear spins, electromagnetic moments and changes in the nuclear root-mean-square charge radii \cite{Yan23,block2021recent}. These observables are key for understanding the atomic nucleus and guiding developments of nuclear theory \cite{Gar16,Mil19,Gor19,Gro20,Rep21,Kos22,Ver22}. In order to extract nuclear properties from atomic spectra, high experimental sensitivity and precision are critical. 
Laser cooling and trapping techniques represent some of the most precise experimental methods \cite{Man19,Hur22}, but they are not universally applicable to all elements of the periodic table, and cannot be employed directly to study short-lived nuclei, typically produced at high temperatures, high energies ($>$ 30 keV), and with sub-second lifetimes \cite{Yan23}. 

A highly successful approach to overcome these challenges is the collinear laser spectroscopy technique applied on bunched ion beams \cite{Neu17}. Unstable isotopes, typically produced from nuclear reactions, can be mass separated as ions, trapped and then cooled in gas-filled radio-frequency ion traps \cite{Neu17,Min13}. Buffer gas collisions are then used to reduce the temperature of the initial beam down to the temperature of the gas. These methods have allowed high-precision measurements of a wide range of radioactive nuclei \cite{Yan23}, and have recently been extended to study radioactive molecules \cite{garcia2020spectroscopy,udrescu2021isotope}. However, the lifetime of the systems that can be studied with such experiments is limited by the cooling and trapping time, which are typically on the order of tens or hundreds of milliseconds \cite{Neu17}.  Moreover, trapping becomes impractical when large amounts of contaminants are present, overfilling the trap and preventing the capture of the ion of interest. New generation radioactive beam facilities, such as the Facility for Rare Isotopes Beams (FRIB) \cite{glasmacher2017facility} in the US and RIKEN in Japan \cite{yano2007radioactive}, are already producing isotopes at the extreme of stability, but due to their short lifetimes ($<$ 5 ms) they cannot be studied with the current laser spectroscopy techniques. Therefore, new spectroscopy methods that could be directly applied to fast, hot, short-lived isotopes need to be developed \cite{ruiz2018high,Jov22}.   

\raggedbottom

\begin{figure*}[t]
\includegraphics[width=2.05\columnwidth,height=.28\textheight]{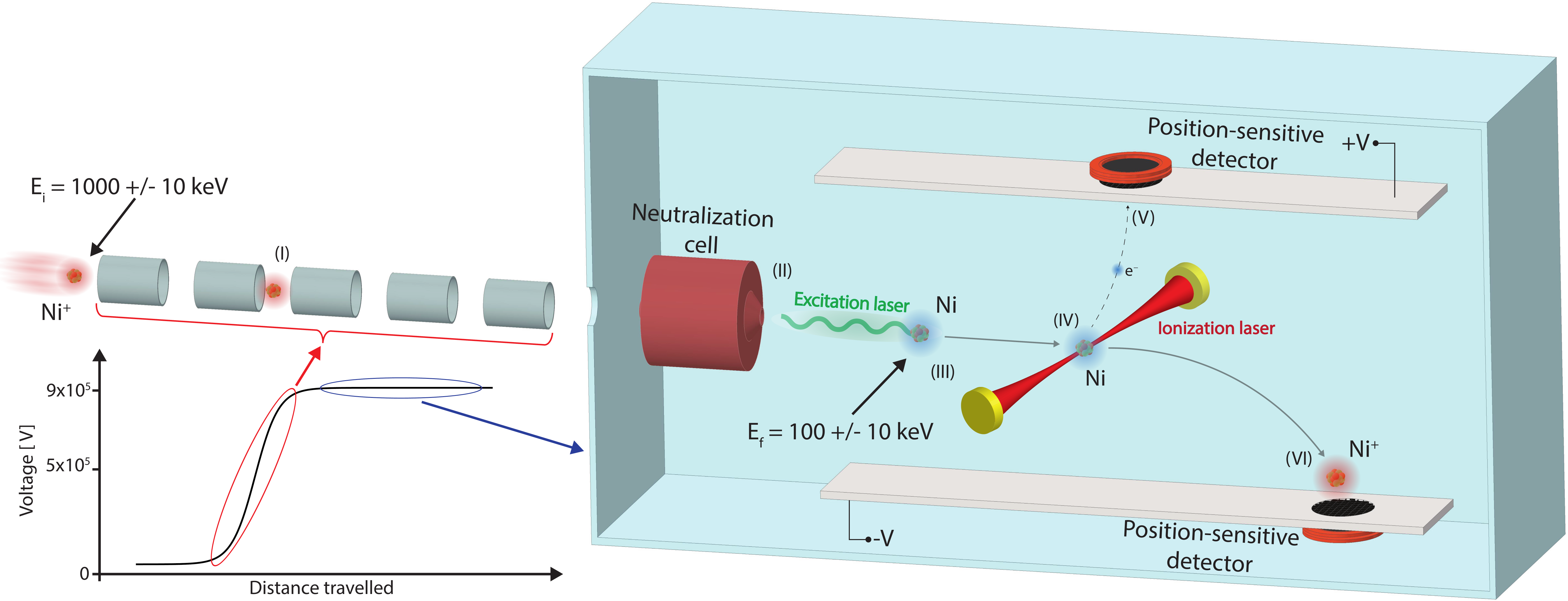}
\caption{Illustration of the experimental setup. Highly energetic ions, $E_i\approx 1$ MeV $\pm 10$ keV, are guided towards a system of electrostatic lenses where their energy is reduced (I) to $E_f \approx 100 \pm 10$ keV. A possible, configuration of voltages as a function of the distance travelled by the ions is shown in the bottom left corner. After that, they enter the interaction region chamber (light blue), kept at a high voltage ($\sim 9\times 10^5$ V), pass through a neutralization cell (II) and are then excited to a higher-lying electronic state by a collinear laser (III). The excited atoms are then ionized by a standing-wave laser inside an optical cavity (IV). The resulting electrons are detected by a position-sensitive detector located above the ionization point (V), while the ions continue their trajectories in the electric field produced in between two parallel plates, until they reach a second position-sensitive detector (VI). For smaller initial energies, $E_i \lesssim 100$ keV, the deceleration lenses, as well as the high voltage applied to the interaction chamber can be removed from the setup.}
\label{experimental_setup}
\end{figure*}

Here, we propose a simple, versatile method for performing high-precision laser spectroscopy studies of fast atomic beams, in a way that allows the energy of the atoms of interest to be measured in-flight on an event-by-event basis. By performing coincidence measurements of the resonant ions and ejected electrons produced during a resonant laser ionization process, the vector velocity of each initial atom can be reconstructed, enabling accurate measurements of the transitions of interest. This setup would enable precision laser spectroscopy measurements directly on isotopes produced by in-flight reactions, with energy spreads as large as 10 keV and lifetimes bellow one millisecond. We demonstrate that precision measurements can be achieved even for arbitrary energy distributions of the initial particles, without the need for cooling mechanisms. 


\section{Experimental Approach}
A schematic of our proposed method is shown in Fig. \ref{experimental_setup}. To illustrate the capabilities of the approach, an initial high energy ion beam with an average energy $E_i = 1$ MeV and an energy spread of  $\Delta E= 10$ keV was assumed. These parameters are similar to the properties of ions produced by in-flight nuclear reactions \cite{glasmacher2017facility,yano2007radioactive}. The ions are decelerated using a series of electrostatic lenses to $E_f = 100 \pm 10$ keV. 
The ions then enter the interaction chamber, kept at high potential ($\sim 9\times 10^5$ V for the considered setup) and are neutralized in a charge exchange cell, filled with a vapor gas (e.g. Na or K vapour \cite{Ver19}). After the non-neutralized ions are removed, the remaining atoms are overlapped collinearly with a continuous-wave laser, such that they are resonantly excited to a particular electronic state. After this, the atoms can be ionized by a different continuous-wave laser beam, perpendicular to the atoms' propagation direction. As this second step can be a non-resonant process, power densities on the order of tens to thousands of kW/cm$^2$ can be required for efficient ionization \cite{demtroder2015laser}. This can be achieved by using a standing-wave laser built inside an optical cavity \cite{carstens2013large}.  The ionization step takes place within the electric field created between two parallel plates. The frequency of the ionization laser can be selected to either directly ionize the atom to the continuum or to excite it to a Rydberg state, from where it can be ionized by the electric field. A position-sensitive detector, located right above the interaction region, can be used to detect individual electrons produced during the ionization process \cite{dorner2000cold,ullrich2003recoil,jiang2022reaction}. This will allow the extraction of the atom's initial position and time at the moment of the ionization. After that, the ion will move in the existing electric field until it reaches a second  position-sensitive detector, such that the ion's final position and time of flight can also be recorded. The location of the two detectors can be chosen such that virtually all the produced electrons and ions are detected (up to the intrinsic efficiency of the detectors, which can be above 90 \%). Using the initial and final position of the ions, as well as their time of flight, the initial ion velocity can be inferred as described below. The same technique can also be used for studying ions produced with low energies, $< 60$ keV, as it is the case at Isotope Separation On-Line (ISOL) facilities \cite{dilling2014isac,borge2017isolde,moore2014igisol}, and for ions produced by arbitrary sources. 


\begin{figure}[]
\includegraphics[width=1\columnwidth,height=0.29\textheight]{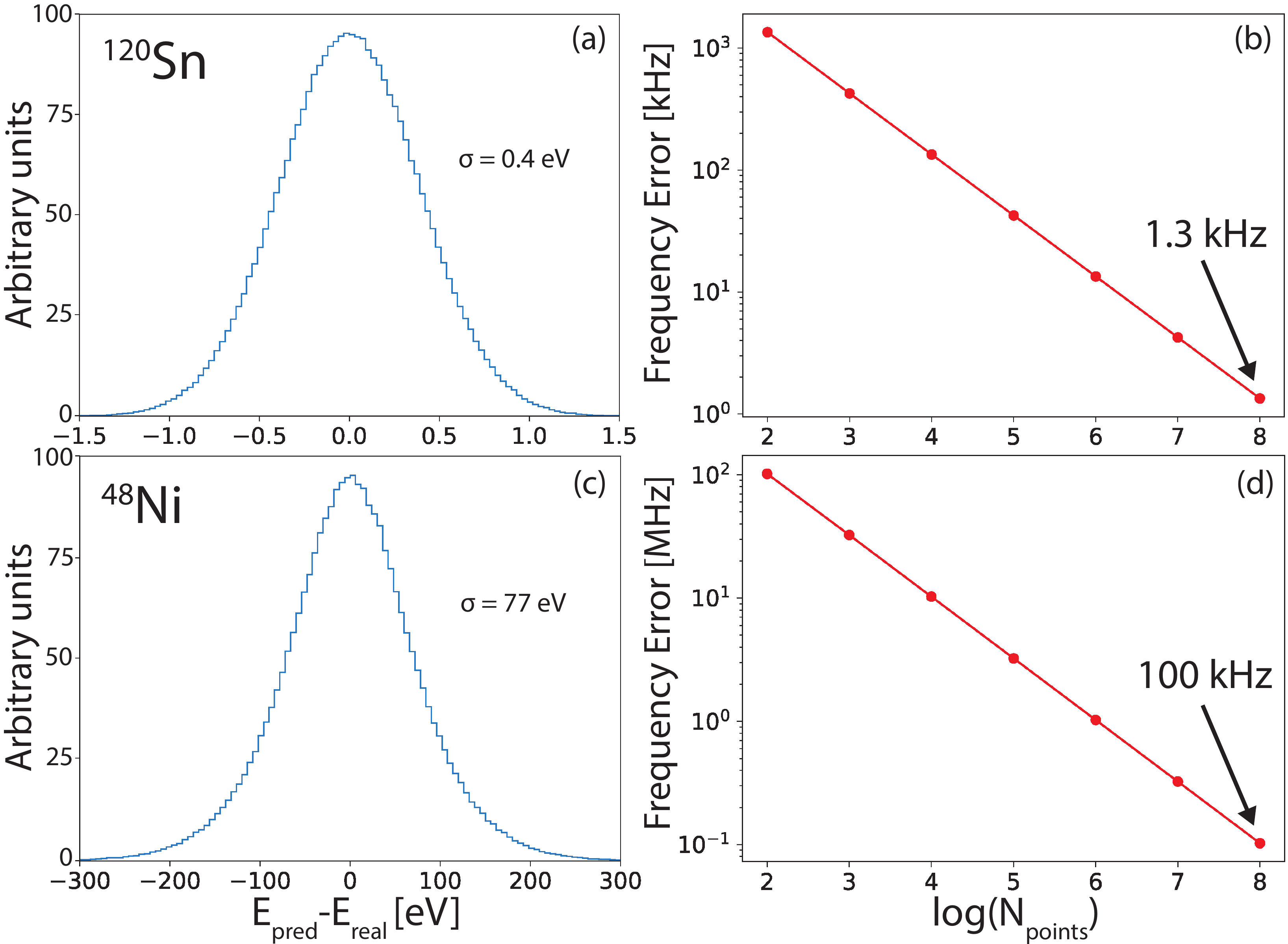}
\caption{Results of the energy and frequency reconstruction. The predicted energy error for $^{120}$Sn over all events is normally distributed with a standard deviation of only $0.4$ eV (a). This corresponds to an event-by-event reconstructed rest frame transition frequency with an error around $1$ MHz when 100 target atoms events are detected, which is further decreased to only $1.3$ kHz for $10^8$ events (b). For $^{48}$Ni, an energy uncertainty of $77$ eV is obtained (c), leading to a reconstructed rest frame transition frequency uncertainty at the MHz level, for $10^6$ events (d).}
\label{ideal_case}
\end{figure}

In the ideal case of a charged particle moving in an uniform field created by two parallel plates, one can analytically compute the particle trajectory and hence extract its initial velocity. In reality, different experimental uncertainties need to be accounted for, such as edge effects due to the finite size of the electrodes or uncertainties in the applied potentials and geometry of the experimental array. The effect of these uncertainties can be overcome by using a reference atom with a well-known electronic transition. To provide a realistic example, $^{40}$Ca and its $^1S_0 \left(4\mathrm{s}^2\right)$  $\rightarrow$ $^3P_1 \left(4\mathrm{s}4\mathrm{p}\right)$ transition at $657$ nm \cite{lurio1964lifetime,engleman1999high} can be used as a reference. For each ionization event, given the known frequency of the excitation step, $\nu_{laser}$, as well as the real frequency in the atom's rest-frame, $\nu_{atom}$, the ion velocity, $v$, and thus energy, $E$ can be extracted using the Doppler correction formula. The uncertainty on the measured energy for each event is given by $dE = m_r v c \left(1-\frac{v}{c}\right)\frac{d\nu_{atom}}{\nu_{atom}}$, where $d\nu_{atom}$ is the linewidth of the transition, $m_r$ is the atom's mass and $c$ is the speed of light. The experiment will then be repeated with the atom of interest, labelled as the target atom, recording its initial and final position, time of flight, and the laser frequency measured during an ionization event.

For a system of non-relativistic particles experiencing only electrostatic fields, two ions with different masses but the same charge state follow identical trajectories. Moreover, the time of flight between two given points is related simply by the square root of the mass ratios of the two species, $R \equiv \sqrt{m_r/m_t}$, where $m_t$ is the mass of the target atom.
Thus, the energy of the atom of interest can be obtained by comparing its time of flight, initial and final positions with those of the reference atom. By predicting the energy of each individual target atom, while knowing the frequency of the laser used during an ionization event, the transition frequency in the atom's rest frame can be obtained on an event-by-event basis by using a neural network (NN).

As the parameter space of the two species is almost identical, neural networks are ideally suited for the task, given their well-known power of interpolation. The main drawback of standard, feed-forward NN is that there is no statistically consistent way of estimating the uncertainties associated with the predicted values, which is critical for high-precision experiments. To overcome this challenge, a Mixture Density Network (MDN) can be used \cite{bishop1994mixture}. Similar to regular NNs, MDNs take a vector as input, which in this case is $\mathbf{x} = (x_i,x_f,t)$ for each individual ion, which is then passed through one or more hidden layers. However, unlike standard NNs, where the output is a deterministic function of the input, the output of an MDN is represented by a mixture of Gaussian functions:

\begin{equation}
    \mathbf{y}(\mathbf{x}) = \sum_{i=1}^N\alpha_i(\mathbf{x})\mathcal{N}\left(\mu_i(\mathbf{x}),\sigma_i(\mathbf{x})\right),
\end{equation}
where $\mu_i(\mathbf{x})$, $\sigma_i(\mathbf{x})$ and $\alpha_i(\mathbf{x})$ are functions learnt during the NN training, and N is the number of Gaussian components. In addition, the loss function of an MDN is a log-likelihood loss:

\begin{equation}
    \mathcal{L} = -\ln\left(\sum_{i=1}^N\frac{\alpha_i(\mathbf{x})}{\left(2\pi\right)^{m/2}\sigma_i(\mathbf{x})}\exp{\left[-\frac{|\mathbf{E}-\mu(\mathbf{x})|^2}{2\sigma_i(\mathbf{x})^2}\right]}\right),
\end{equation}
where $\mathbf{E}$ is the vector of energies of the training data. Thus, after training the network using the reference atom measurements, the energy value and the associated uncertainty can be predicted for each event measured for the atom of interest.

\section{Results and discussion}

To illustrate the overarching capabilities of our approach we have selected two different isotopes, $^{48}$Ni and $^{120}$Sn, as the atoms of interest using their $^3F_3 \left(4\mathrm{s}^2\right)$  $\rightarrow$ $^5P_2 \left(4\mathrm{s}4\mathrm{p}^2\right)$ transition at $255$ nm \cite{huber1980oscillator} and $^3P_0 \left(5\mathrm{s}^25\mathrm{p}^2\right)$  $\rightarrow$ $^1S_0 \left(5\mathrm{s}^25\mathrm{p}^2\right)$ transition at $583$ nm \cite{Lei22}, respectively. Precision isotope shift measurements for the isotopic chains of Ni and Sn are of marked interested for nuclear structure \cite{gus20,Rod20,Mal22,Gor19}, nuclear matter \cite{sky21,Jun21}, and new physics searches \cite{Rei20,mul21,Lei22}. These two elements have not been laser cooled yet, consequently,  achieving precision measurements with the current laser spectroscopy techniques is particularly challenging. 

With both neutron and proton magic numbers at $N=20$ and $Z=28$, respectively, $^{48}$Ni is  of marked interest for nuclear structure and the study of exotic nuclear phenomena at the edge of stability \cite{Bla00,Gri00}. Although this isotope can be produced at the current radioactive beam facilities, its short lifetime ($\tau$=2.1 ms) prevents its study using of the current laser spectroscopy techniques.  $^{48}$Ni is the mirror nucleus of the stable doubly magic $^{48}$Ca, hence a measurement of the nuclear charge radii difference between these two isotopes can be highly sensitive to constrain parameters of the equation of state of nuclear matter \cite{sky21}. On the other hand, Sn is the element with largest number of stable isotopes (ten). Precision isotope shift measurements over this long isotope chain could provide complementary studies to constrain the possible existence of new fundamental forces and particles \cite{Lei22}. Having a proton magic number, $Z=50$, Sn isotopes posse a relatively simple nuclear structure and thus exhibit reduced sensitivity to Standard Model effects, such as nuclear deformation \cite{Lei22,Yor20}, facilitating a clear identification of any new physics signals.

Numerical simulations of electric fields and ion beam trajectories were performed using the software SIMION \cite{simion}. In order to prove our ability to properly reconstruct the correct rest frame transition frequencies of interest, the ions and electrons are produced in between the two parallel plates, just below a position-sensitive detector. The ions' initial position is assumed to be distributed according to a 3D Gaussian with a standard deviation of $1$ mm. This is a reasonable value given the laser beam diameters that can be achieved in high-power optical cavities \cite{carstens2013large}. The simulations were performed with a potential difference of $6$ kV and $20$ kV between the two plates, located $20$ cm apart, for the $^{120}$Sn and $^{48}$Ni case, respectively. The electrons are assumed to be produced with nearly zero kinetic energy, which can be achieved by setting the ionization laser close to the IP of the atom or by exciting the atom to a Rydberg state and then performing field ionization. The time-of-flight of the electrons was calculated, resulting in a distribution with a standard deviation of less than $30$ ps, which is below the resolution of typical position-sensitive detectors (which we assume in this paper to be $50$ ps). This uncertainty was added in quadrature to the time-of-flight uncertainty due to detectors' resolution, $dt$. The uncertainties on the initial, $dx_i$ and final, $dx_f$ position are given by the detector's resolution, which is assumed to be $10$ $\mu$m. The initial energy of the ions, both for the reference and target species, is assumed to be a Gaussian of mean $1000$ eV and a standard deviation of $100$ eV for the $^{120}$Sn simulations. A mean of $100$ keV and standard deviation of $10$ keV was assumed for the $^{48}$Ni simulations (which is the expected energy after the deceleration step). For the reference atom, for each event, a Gaussian noise of mean zero and standard deviation given by the above-mentioned uncertainties ($dx_i$, $dx_f$, $dt$, $dE$) was added to the values of the initial and final position, time-of-flight, and initial ion energy, which are then recorded. The same steps were followed for the target atom, except that the initial energy information was assumed to be unknown. 

For our numerical simulations, for each of the two considered cases, $2\times 10^8$ ions were generated, corresponding to the reference atom. Half of them were used for training and the other half for validation, allowing us to optimize the hyperparameters of the MDN. The MDN used in this work, implemented in PyTorch \cite{NEURIPS2019_9015}, had one hidden layer of 10 nodes, an ELU activation function and one Gaussian component. We trained it for 6000 epochs with a starting learning rate of $10^{-2}$, which was reduced by a factor of 10 every 1000 epochs. We used the Adam optimizer with a batch size of 1024. Using the trained neural network, we predicted the energy and associated rest frame transition frequency uncertainty for each event of the target atom. Figure \ref{ideal_case} show the obtained results. For the simulations involving $^{120}$Sn, Fig. \ref{ideal_case} (a) displays a histogram of the MDN's prediction error defined as $E_{pred}-E_{real}$ in eV, having a standard deviation of $0.4$ eV. This is a significant uncertainty reduction relative to the initial spread of $100$ eV. For $^{48}$Ni, the energy uncertainty was reduced from $10$ keV to only $77$ eV (Fig. \ref{ideal_case} (c)). In both cases, $\approx 95\%$ of the MDN's predictions are within two standard deviations from the true energy value, proving the reliability of the MDN's estimation of individual energies and associated uncertainties. 

\begin{figure}[t]
\includegraphics[width=\columnwidth,height=.3\textheight]{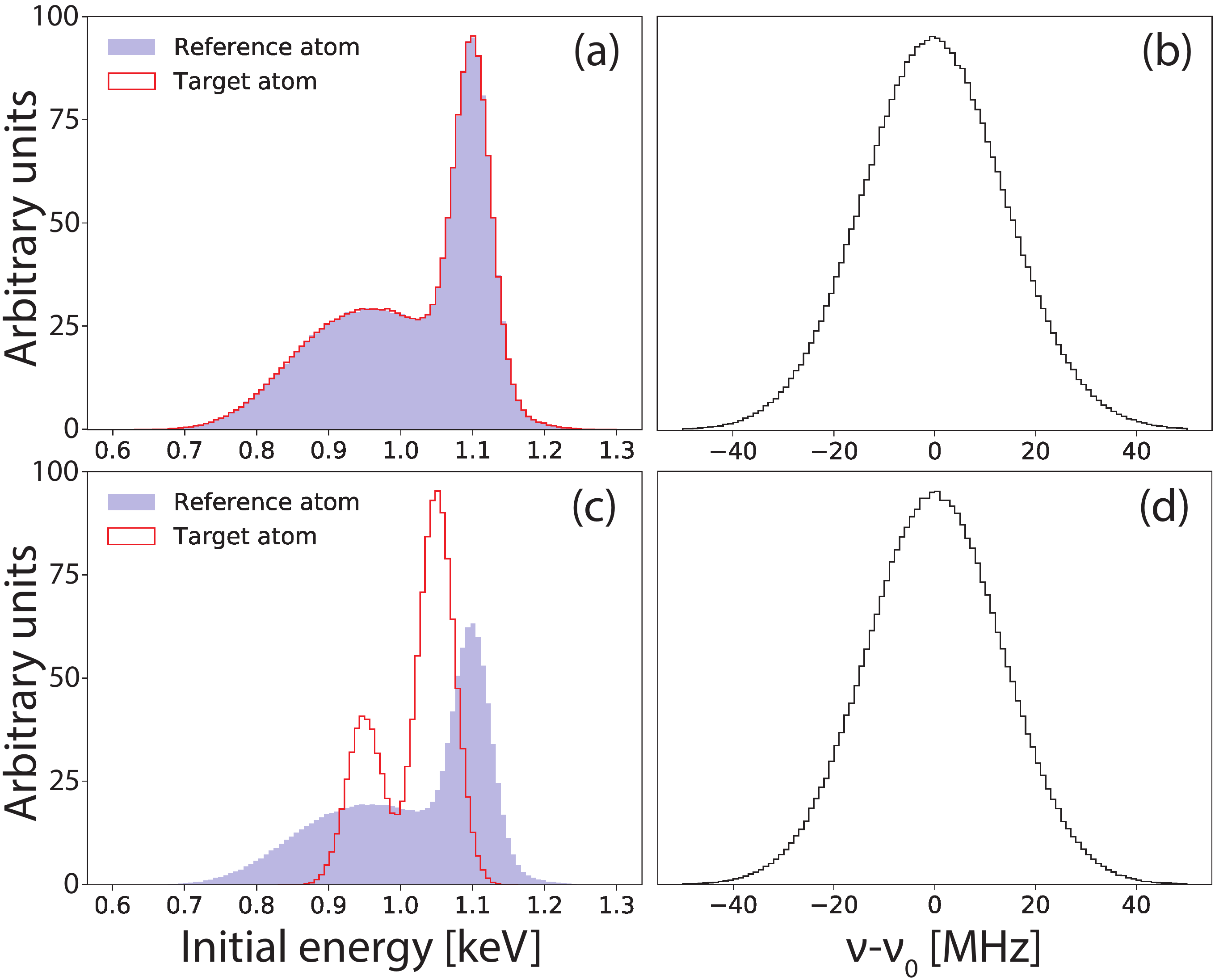}
\caption{Results of the frequency extraction
for ions with non-Gaussian distributed initial energies, for the $^{120}$Sn case. The energy histograms when the reference and target atom have the same/different initial energy distributions are shown in (a)/(c), while the associated reconstructed rest-frame frequency of the target atom using our proposed method, relative to the true rest-frame frequency, $\nu_0$, is shown in (b)/(d).}
\label{weird_initial}
\end{figure}

Using the predicted mean and standard deviation of the energy, as well as the laser frequency used when an event was observed, we can calculate the transition frequency in the rest frame of the target atom, together with its associated uncertainty on an event-by-event basis. In our simulation this is done as follows. For each event we sample the true rest-frame frequency using a Lorentzian distribution with mean and linewidth given by the transition of interest mentioned above. We Doppler shift it to the lab frame using the real energy of the ion, taken from SIMION, thus obtaining the laser frequency at which the given event was observed. Finally, we Doppler correct back to the ion's rest frame using the MDN predicted energy. The final value of the transition, obtained from N measured events, will be given by: $\nu_{pred} = \sum\limits_i^N{\frac{\nu_i}{d\nu_i^2}} \bigg/\sum\limits_i^N{\frac{1}{d\nu_i^2}} $ and its associated uncertainty:  $d\nu_{pred} = 1\bigg/\sum\limits_i^N{\frac{1}{d\nu_i^2}}$, where $\nu_i$ and $d\nu_i$ are the predicted frequency and uncertainty, respectively, for the i-th target atom event. In Fig. \ref{ideal_case} (b) and (d) we show the error in the calculated frequency as a function of the number of target atom events, for $^{120}$Sn and $^{48}$Ni, respectively. In the former case, we can correctly predict the true value of the rest-frame transition frequency with an uncertainty of about $1$ MHz with as little as $100$ events and reach a  precision of $1$ kHz when $10^8$ events of the target atom are measured. In the latter case, an uncertainty at the MHz level is achieved for only $10^6$ events, which is suitable for extracting electromagnetic nuclear properties of radioactive elements with lifetimes in the millisecond range, as is the case for $^{48}$Ni \cite{Yan23}. We emphasize that performing measurements on such short lived elements is out of the reach of the current experimental techniques. Note that the increased uncertainty in the second case is due to the larger beam energy and rest frame transition frequency, compared to the $^{120}$Sn case.

To study possible systematic uncertainties, numerical simulations under the following experimental conditions were performed: (i) one of the plates was tilted with respect to the beam propagation direction at different angles between $0$ and $5$ degrees, (ii) both plates were simultaneously tilted with respect to the horizontal at angles between 0 and 2 degrees, (iii) up to 10$\%$ relative uncertainties on the voltage applied to the two plates as well as on the distance between the plates were assumed, (iv) up to 5 millimeters of uncertainty on the vertical location of the ionization was assumed, (v) an additional transverse energy component of up to $1$ keV to the ions was introduced. These values are significantly larger than one would expect from a properly implemented experimental setup. However, these uncertainties do not have a significant effect on the energy reconstruction of events. This is expected, as described above, given that during training, the MDN is able to learn all these variations of the experimental parameters which are encoded in the trajectories of the reference atoms and account for them in the predictions of the energies of the target atom.

To further explore the robustness of our approach in the case where the initial energy is not symmetrically distributed, we performed simulations assuming an arbitrary energy distribution, as shown in Fig. \ref{weird_initial} (a). This is particularly interesting as it indicates that our proposed scheme can be applied universally, for atoms and/or ions produced by diverse physical processes. For such a situation, simply applying a regular Doppler shift to the entire atomic ensemble, assuming a fixed acceleration voltage, would lead to distorted lineshapes and add significant systematic uncertainties to the value of the extracted transition. However, as our method can recover the energy and hence the rest-frame frequency of individual atoms, using only the recorded parameters, it allows the scheme to be independent of the original energy distribution and produce consistent results for arbitrary distributions. This is shown in Fig. \ref{weird_initial} (b) for the case of $^{120}$Sn, where the distribution of the reconstructed rest-frame transition frequency has a well-behaved shape, from which the correct value of the frequency can be extracted with uncertainties at the $100$ kHz level, with only $10^{4}$  events.

Finally, we explore the situation in which the reference and target atoms have different initial energy distributions (Fig. \ref{weird_initial} (c)). This is of interest when the production mechanism is a particularly violent one, as is often the case at radioactive beam facilities. The results obtained in this case are shown in Fig. \ref{weird_initial} (d). Again, we can recover the correct rest-frame transition frequency with similar uncertainties as described above, using arbitrary distributions for the initial energy. Similar results were obtained for $^{48}$Ni atoms with initial energies on the order of $\sim 100$ keV, where rest frame transition frequencies with uncertainties of $10$ MHz can be obtained for only $10^4$ events. 

The main conditions that need to be fulfilled for the energy reconstruction to be successful in our setup are: (i) the parameter space of the target atom needs to be contained within the parameter space of the reference one and (ii) enough events need to be recorded during the reference atom measurement, for any region of interest in the parameter space. The first condition should be fulfilled by the geometric constraints of the setup i.e., the electron and ion should hit the position-sensitive detector for both ions. Moreover, one can easily compare the spatial and temporal information of the two species and confirm the validity of this assumption. The second condition depends only on the statistics accumulated while measuring the reference atom, which can be easily increased by several orders of magnitude if needed.

\section{Conclusion}

We proposed a simple, versatile and powerful experimental technique, suitable for performing high-resolution spectroscopy on fast, hot, short-lived isotopes produced with arbitrary energy distributions and large energy spreads. Using the temporal and spatial information of ions and corresponding electrons produced in a laser ionization process, the atoms's initial vector velocity can be reconstructed on an event-by-event basis, enabling precision measurements of the rest-frame transition frequency of the species of interest.  

The described method is very efficient in terms of computing time, as the MDN needs to be trained initially,
only once for a given energy range, using a reference
atom. After that, predictions can be made for any other atom and transitions of interest. By increasing the number of events for the atom of interest, $N_t$, the uncertainty on the measured transition frequency, which scales as $\frac{1}{\sqrt{N_t}}$, can be further reduced. This can also be achieved by using detectors with better spatial and temporal resolutions, compared to the ones considered here. Possible NN prediction biases can be estimated using the validation data set of the reference atom. The MDN predictions could potentially be further improved by using different learning rate schedules or trying more complex architectures (in terms of layers, nodes per layers or number of Gaussian components), compared to the one presented in this work. The performance of other types of NN, able to estimate uncertainties for the predicted results, such as Bayesian NN \cite{neal2012bayesian}, can also be investigated. Non-NN based methods could also offer a possible pathway for estimating the energies of interest. For example, for each target event, the energy and the associated uncertainty could be estimated in terms of the mean and standard deviation of the energy of the closest reference events in parameter space, where the number of such events would be optimized using the validation data.

The ability to perform precision measurements on fast beams with large energy spreads opens the way to studying short-lived isotopes at the extreme of existence, with lifetimes on the order of milliseconds or less. Such species can already be produced  at different radioactive beam facilities worldwide, but their laser spectroscopic investigation is currently impossible with current techniques. The proposed approach also opens the possibility to study isomeric states with sub-millisecond lifetimes that can be produced in-flight by nuclear reactions. Such short-lived species are critical for exploring the limits of nuclear existence and guiding our fundamental understanding of atomic nuclei \cite{Yan23,block2021recent}.

\section*{A\titlelowercase{cknowledgments}}

\begin{acknowledgments}
The authors thank the discussions with Shane G. Wilkins, Fabian Pastrana and Maria Paula Montes. This work was supported by the Office of Nuclear Physics, U.S. Department of Energy, under grants DE-SC0021176 and DE-SC0021179; the Jarve Seed fund; D.A. Torres thanks the Laboratory of Nuclear Science at MIT for their hospitality during the sabbatical year when this work was performed, and the Universidad Nacional de Colombia for the support under grants 51111 and 51119. 
\end{acknowledgments}

\bibliography{apssamp}

\end{document}